\documentstyle[aps,pra,epsf,multicol]{revtex}
\begin{document}
\draft
\newcommand{\ket}[1]{\left | #1 \right \rangle}
\newcommand{\bra}[1]{\left \langle #1 \right |}

\title{Quantum Stochastic Resonance in Electron Shelving}
\author{S.F. Huelga$^{1,2}$ and M.B. Plenio$^{1}$}
\address{${(1)}$ Optics Section, The Blackett Laboratory, Imperial College, London SW7 2BW,
England\\ ${(2)}$ Department of Physics, University of Hertfordshire, Hatfield,
England.}

\maketitle
\begin{abstract}
Stochastic resonance shows that under some circumstances noise can enhance the
response of a system to a periodic force. While this effect has been
extensively investigated theoretically and demonstrated experimentally in
classical systems, there is complete lack of experimental evidence within the
purely quantum mechanical domain. Here we demonstrate that stochastic resonance
can be exhibited in a single ion and would be experimentally observable using
well mastered experimental techniques. We discuss the use of this scheme for
the detection of the frequency difference of two lasers to demonstrate that
stochastic resonance may have applications in precision measurements at the
quantum limit.
\end{abstract}

\pacs{Pacs No: 03.65.Sq, 42.50.Lc, 05.45.+b}

\begin{multicols}{2}
Imagine that you are set the task of detecting a very weak periodic signal by
means of its interaction with a suitable probe system. Given that the signal is
{\em weak}, one intuitively may think that the optimal experimental set up
should minimize any other interaction that the probe may undergo. However, this
is not always the case and there are situations where noise can indeed play a
constructive role in a high sensitivity detection. A clear illustration of this
fact is provided by the phenomenon of stochastic resonance (SR) \cite{benzi},
where the response of a nonlinear system to external periodic driving is
enhanced in the presence of noise. Typically \cite{rubi2}, the signal-to-noise
ratio (SNR) increases monotonically up to a maximum for certain optimal noise
intensity, and then decreases gradually as randomization dominates over the
cooperative effect between the coherent driving and the stochastic forces. The
simplest system for describing the appearance of SR consists of a particle in a
bistable potential subject to both thermal noise and a periodic forcing
\cite{wiesenfeld}. However, many other scenarios have been proposed and recent
research has shown that SR may also be observed in some monostable systems
\cite{rubi1}. Experimental research has confirmed that the phenomenon, if at
first sight counter-intuitive, is rather ubiquitous. Since the first
demonstration in a Schmitt trigger circuit \cite{trigger}, SR has been observed
in a wide variety of physical systems, ranging from ring lasers to neuronal
cells (see \cite{Haenggi} for a detailed review). Moreover, there is
experimental evidence that certain complex living systems (such as crickets)
make use of SR to improve the sensitivity of their sensory organs
\cite{bichos}.

\begin{figure}
\begin{center}
\leavevmode \epsfxsize=7cm \epsfbox{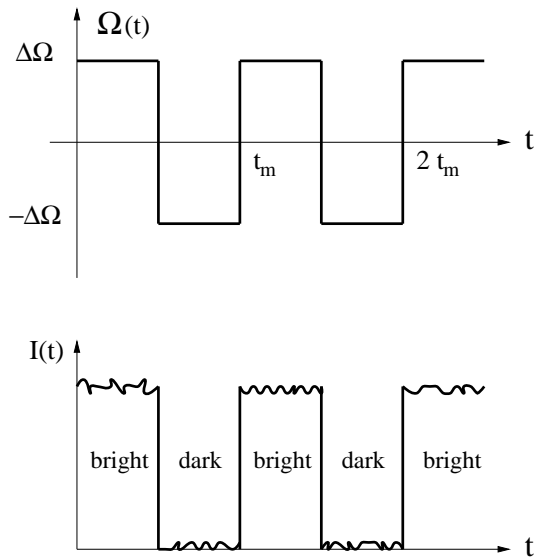} \vspace*{.4cm}
\caption{\narrowtext By means of suitably tuning the broad band fields driving
the $1\leftrightarrow 3$ and $1\leftrightarrow 4$ transitions, the system may
exhibit a bistable dynamics of bright and dark periods strongly synchronized
with the modulation of the coherent driving.} \label{saltos}
\end{center}
\end{figure}

Recently the concept of SR has been extended to the quantum domain
\cite{Loefstedt,Reale,Grifoni}. However, experimental verifications at the
level of individual quantum systems are difficult (see e.g.
\cite{Buchleitner}) and it would be of great interest to find
feasible experimental scenarios that allow the investigation of stochastic
resonance in the quantum regime. In this letter we show that SR can be
demonstrated in a conceptually simple truly microscopic quantum optical system.
We first analyze the proposed system {\em qualitatively}, highlighting the key
ideas behind our proposal and allowing for an intuitive understanding as to why
SR arises in this scenario. Following this, we present a {\em quantitative}
analysis of the phenomenon by means of exact numerical computation of the
frequency response of the probe. We demonstrate that the SNR at the driving
frequency is maximized at a certain noise level, an unambiguous signature of
the occurrence of SR. Furthermore, to illustrate an application of SR in our
proposed system, we discuss and simulate the noise-assisted precision
measurement of the frequency difference of two coherent fields.
\begin{figure}
\begin{center}
\leavevmode \epsfxsize=8cm \epsfbox{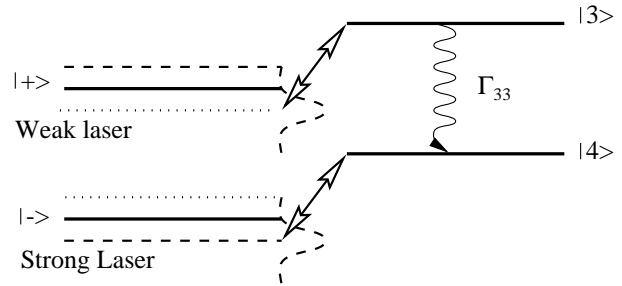} \vspace*{.4cm}
\caption{\narrowtext Four level system under coherent modulated driving with
time-dependent Rabi frequency $\Omega$ giving rise to dressed states $|\pm\rangle$.
Dashed (dotted) line gives position of dressed states for larger (smaller) value
of $\Omega$. Transitions $-\leftrightarrow 3$ and $+\leftrightarrow 4$ are driven
by broad band fields. The frequency distribution of the noisy fields is represented
by the dashed curves. When the central frequency of each incoherent driving is tuned
appropriately, the interaction can be made resonant with either the dressed level
$\ket{+}$ or $\ket{-}$ depending on the value of $\Omega$ given by the modulation
cycle (See text for details).} \label{setup}
\end{center}
\end{figure}
\vspace*{-0.2cm} The scheme we present here is in principle easy to implement,
as it relies entirely on techniques have been employed by experimentalists
for more than 10 years. In Fig. \ref{setup} the system under consideration
and the applied driving fields are shown. The probe consists of a four level
atomic system subject to both coherent and incoherent radiation. The
$1\leftrightarrow 2$ transition is driven by a resonant modulated coherent
laser field of Rabi frequency $\Omega(t)$ while the $1\leftrightarrow 3$ and
$2\leftrightarrow 4$ transitions are driven by noisy fields with effective pump
rates $W_{33}$ and $W_{44}$ respectively. The atomic level $2$ can decay with a
rate $2\Gamma_{22}$ and it is this radiation that will be detected. We assume
that level $4$ is metastable and we neglect its spontaneous decay rate in the
following. In addition, we assume that level $3$ can only decay, at a rate
$2\Gamma_{33}$, into level $4$. The fact that the coherent driving is modulated
implies that the Rabi frequency and therefore the energy separation $\hbar
\Omega$ between the dressed levels $\ket{\pm} =1/\sqrt{2} (\ket{1} \pm
\ket{2})$ are time dependent. Let us consider the simplest case where the
modulated driving can be described by a step function, as illustrated in Fig.
\ref{saltos}. Then the time dependence of the coherent Rabi frequency is given
by
\begin{equation}
    \Omega(t)= \Omega \pm \Delta \Omega,
    \label{eq1}
\end{equation}
where $\Delta\Omega \ll \Omega$ (weak modulation), and where the $(+)$ sign
holds for $t \in [k t_m, (k+1/2)t_m]$, k is a positive integer, and the $(-)$
sign corresponds to the remaining part of the modulation cycle, $t_m$ denoting
the corresponding modulation period. Under these conditions, the relative position
of the dressed states alternates between two possible configurations, as illustrated
by the dashed (dotted) lines in Fig. \ref{setup}. For the sake of clarity, we will refer to
these two possible situations as {\em strong laser} (i. e. larger Rabi frequency;
larger energy separation between dressed levels) and {\em weak laser} (i. e. smaller
Rabi frequency; smaller energy separation between dressed levels) regimes.
We
will now show in a qualitative way that the system we have described above may
exhibit, when the broad band fields are suitably tuned, a dynamics of extended
bright and dark periods in the resonance fluorescence intensity which are strongly
synchronized with the modulated coherent driving field, as depicted in Fig. \ref{saltos}.
However, synchronization is not the only signature of SR and we will show quantitatively
later on that indeed the proposed scheme fulfills the necessary requirements
for exhibiting SR.

\begin{figure}[t]
\begin{center}
\leavevmode \epsfxsize=5.5cm \epsfbox{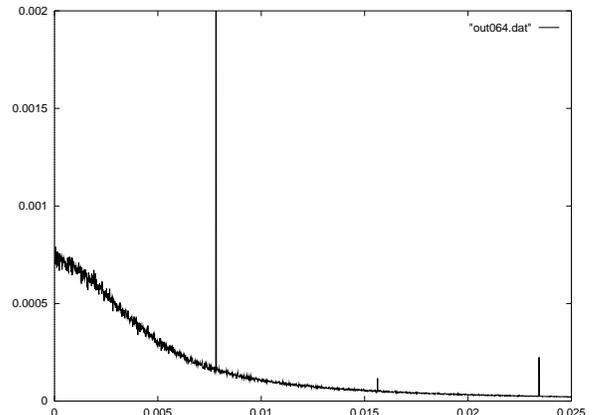} \caption{\narrowtext
\label{switch} Numerical simulation of the power spectrum for a modulation as
in Eq. (\ref{eq1}). The other parameters are $\Omega=50
\Gamma_{22},\Delta\Omega=10 \Gamma_{22}$, $\Gamma_{33}=\Gamma_{22}$,
$W_{33,weak}=0.0128$, $W_{33,strong}=W_{33,weak}/10$, and
$W_{44,strong}=W_{33,weak}$, $W_{44,weak}= W_{33,strong}$ and detunings chosen
as in Eqs. (\protect\ref{eq3},\protect\ref{eq4}).
The peaks are delta-functions to within the numerical precision. Harmonics at
even multiples are suppressed because of the symmetry of the modulation. }
\end{center}
\end{figure}

\begin{figure}
\begin{center}
\leavevmode \epsfxsize=6.cm \epsfbox{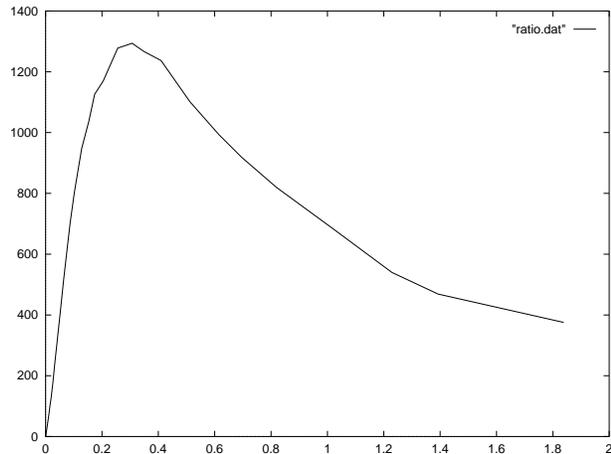} \caption{\narrowtext Output
signal-to-noise (SNR) at the driving frequency as a function of the noise
intensity. The parameters are chosen as in Fig. \protect\ref{switch}.
The sharp rise of the noise intensity to a maximum for an intermediate
value of the noise intensity followed by a slower fall-off towards smaller SNR
is a clear signature of the occurrence of SR.} \label{switch1}
\end{center}
\end{figure}

Let us denote by $\hat \omega_i$, $(i=3,4)$, the central frequency of each
broad band field, while $\omega_{i1}$ refers to the corresponding natural
frequency of the atomic transition involved.  Let us now choose the detunings
$\Delta_{i1}=\hat \omega_i - \omega_{i1}$ as follows
\begin{eqnarray}
    \Delta_{31} &=& \frac{\Omega+\Delta \Omega}{2} \label{eq3}\\
    \Delta_{41}&=&-\,\frac{\Omega-\Delta \Omega}{2} \label{eq4} \; .
\end{eqnarray}
This choice of detunings is schematically shown in Fig. \ref{setup}. When the
coherent driving operates in the {\em weak mode}, that is, when
$\Omega(t)=\Omega-\Delta \Omega$, the noisy field driving the $1\leftrightarrow
3$ transition is resonant with the dressed level $\ket{+}$ while the field
driving the $1\leftrightarrow 4$ is off-resonant. Under these circumstances,
the system is rapidly pumped from $\ket{1}\leftrightarrow \ket{3}
\leftrightarrow \ket{4}$. It remains in level 4 and as a consequence no light
is emitted from the atom, i.e. we are in a dark period, given that level
$\ket{4}$ is metastable. On the other hand, when the coherent driving is
operating in the {\em strong mode}, the noise field driving the
$1\leftrightarrow 4$ becomes resonant with the dressed level $\ket{-}$ while
the driving $1\leftrightarrow 3$ becomes detuned. As a result, the system is
pumped back and forth between level $\ket{4}$ and the dressed level $\ket{-}$,
with photons being emitted at a rate proportional to $\Gamma_{22}/2$.
Therefore, the system is in a bright period, i.e. the atom emits many photons.
It can be understood quite easily how this synchronized dynamics depends on the
noise intensity. If the effective pump rate $W_{ii}$, $(i=3,4)$, is too strong,
e. g. $W_{ii} \gg\Delta \Omega$, the pump rates become insensitive to the value
of $\Delta \Omega$, synchronization is lost and the observed fluorescence would
just show a constant intensity. A lack of synchronization should also be
observable in the opposite regime, where $W_{ii}$ is too weak, e. g. $W_{ii} <
t_m^{-1}$, with the system exhibiting now extended bright and dark periods and
an additional noise background in the fluorescence spectrum. Therefore, there
seems to be an intermediate regime in which synchronization is optimal. We will
demonstrate in the following that this optimal regime can be achieved and how
in fact the system may exhibit stochastic resonance. To clearly identify SR we
have to compute the autocorrelation function of the intensity emitted by the
atom. We simplify it to a binary process, assuming value $1$ if the intensity
exceeds a certain threshold (e.g. $10\%$ of the intensity expected in a bright
period), and $0$ for intensities below the threshold. We then compute the power
spectrum of this process which defines the spectral response of the system to a
periodic perturbation. For that, we will have to compute explicitly the
normalized spectrum and the corresponding SNR at the driving frequency. The
starting point for evaluating these quantities is provided by the system's
master equation, whose relevant terms \cite{aprox}, under exact resonance for
the $1\leftrightarrow 2 $ transition, are as follow:
\begin{eqnarray}
{\dot\rho}_{++} &=& -(2 W_{33}+\frac{\Gamma_{22}}{2})\rho_{++} +
\frac{\Gamma_{22}}{2} \rho_{--} + 2 W_{33} \rho_{33} \label{Eq1}\\
{\dot\rho}_{--} &=& \frac{\Gamma_{22}}{2} \rho{++}
-(2W_{44}+\frac{\Gamma_{22}}{2})\rho_{--} + 2 W_{44} \rho_{44} \label{Eq2}\\
{\dot\rho}_{+-} &=& \Gamma_{22} (\rho_{++}+\rho_{--})\\ \nonumber
&-&(W_{33}+W_{44}-\frac{3 \Gamma_{22}}{2} + i \frac{\Omega}{2})\rho_{+-} -
\frac{\Gamma_{22}}{2} \rho_{-+} \label{plus}\\
{\dot\rho}_{33} &=& 2 W_{33} \rho_{++} -(2 W_{33} - 2 \Gamma_{33}) \rho_{33}\\
{\dot\rho}_{44} &=& = - {\dot\rho}_{++} - {\dot\rho}_{--} - {\dot\rho}_{33},
\;\; {\dot\rho}_{-+} = {\dot\rho}_{+-}^{*}
\end{eqnarray}
The last line arises from the preservation of trace and the hermiticity of the
density operator. It should be stressed that this master equation is valid for
a certain range of parameters in which the broadband assumption for the noise
fields is correct, i.e. we can replace the effect of the noise by a simple pump
rate $W_{ii}$. This approximation is valid if the frequency bandwidth $\Delta
\omega_i$ $(i=3,4)$ of the noise field is larger than the spontaneous decay
rates in the system and the detunings do not greatly exceed the bandwidth of
the noise field. It should also be noted here that we are working in a dressed
state picture, in which the incoherent pump rates are between the dressed level
$|\pm\rangle$ and the level $3$ and $4$. This assumption is only valid if the
Rabi frequency of the driving field on the $1\leftrightarrow 2$ transition is
large compared to the bandwidth of the noise fields. This condition is
intuitively clear, as in that case a noise field can selectively address only
one dressed state. Therefore, our analysis applies provided that the following
inequality holds
\begin{equation}
\Gamma_{ii} \ll \Delta \omega_i \ll \Omega, \,\, (\small i=3,4).
\end{equation}
\normalsize
The master equation contains all the information about the dynamics
of the system; however, if we are mainly interested in the behaviour of a
single quantum system, e.g. a single ion in an ion trap, then it is more
convenient to use the quantum jump approach ( see \cite{Plenio3} and references
therein). The main idea behind the quantum jump approach is to determine the
time evolution of the system under the condition that no photon has been
emitted on the $1\leftrightarrow 2$ transition. This conditional time evolution
is no longer trace preserving and the decreasing trace reflects the
probability that no photon has been emitted in the time interval $[0,t]$. The
conditional time evolution can easily be obtained either directly from the
master equation (removing some terms) or by rederiving it under the constraint
that no photon has been emitted \cite{Plenio3}. The key point is realizing that
in order to obtain the conditional time evolution one has to remove from the
ensemble all those systems that have emitted a photon. This can be done
heuristically by removing the contribution $2\Gamma_{22}\rho_{22}$ from the
time evolution equation for the population of the ground state $\rho_{11}$.
When using dressed states $\ket{\pm}$, the result is that we need to replace
Eqs. (\ref{Eq1},\ref{Eq2},\ref{plus}) by
\begin{eqnarray}
{\dot\rho}_{++} &=& -(2 W_{33}+ \Gamma_{22})\rho_{++} \\ \nonumber &+&
\frac{\Gamma_{22}}{2} \rho_{+-} + \frac{\Gamma_{22}}{2} \rho_{-+}+ 2 W_{33}
\rho_{33} \label{Eq6}\\
{\dot\rho}_{--} &=& -(2 W_{44}+ \Gamma_{22})\rho_{--} \\ \nonumber &+&
\frac{\Gamma_{22}}{2} \rho_{+-} + \frac{\Gamma_{22}}{2} \rho_{-+}+ 2 W_{44}
\rho_{44} \label{Eq7}\\
{\dot\rho}_{+-} &=& \frac{\Gamma_{22}}{2} (\rho_{++}+\rho_{--})\\ \nonumber
&-&(W_{33}+W_{44}+\Gamma_{22} + i \frac{\Omega}{2})\rho_{+-} \label{Eq8}
\end{eqnarray}
Given the conditional time evolution, the evaluation of the normalized
power spectrum is a straightforward task. For the modulation described
in Eq. (\ref{eq1}) numerical results are presented in Fig. \ref{switch}.
This simulation shows the expected behaviour of a resonant-like process,
with a sharp peak (a delta function within numerical precision) at the
frequency of the modulation and subsequent weaker peaks at its odds
harmonics. The peaks at even harmonics are suppressed due to the symmetry
of the modulation. The emergence of delta peaks in the spectrum is a
first indication of SR.\vspace*{-2.7cm}
\vspace*{-0.5cm}
In order to establish unambiguously the occurrence of SR, one has to evaluate
the signal-to-noise ratio. This quantity, defined as the ratio of the spectral
peak to the spectral background at a given frequency, is a measure of the
probe sensitivity to the periodic driving. Fig. \ref{switch1} shows the SNR at the
modulation frequency, i.e. the frequency at which the power spectrum exhibits
a delta peak. As expected, the SNR exhibits a maximum at an optimal noise
pump rate. If the noise intensity is increased beyond this value, the sensitivity of
the probe is diminished. Although we have presented numerical results for a specific
choice of parameters, extensive numerical simulations show that
SR can be observed over a wide range of parameters.
\vspace*{-2.8cm}
\vspace*{-0.4cm} The scheme discussed above may seem academic, but it allows us to
exemplify how stochastic resonance could be observed at the level of a single
quantum system using currently available experimental techniques. Moreover, we
will now show that this phenomenon may find practical applications in precision
measurements. To this end
\vspace*{-2.8cm}
\begin{figure}
\begin{center}
\leavevmode
\epsfxsize=6.cm
\epsfbox{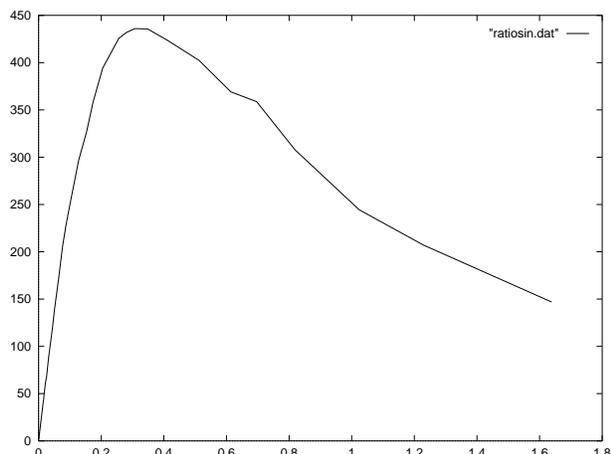}
\caption{\narrowtext \label{switch2}
Output signal-to-noise (SNR) at the driving frequency as a function of the noise
intensity for the two beating lasers driving the $1\leftrightarrow 2$
transition. The parameters are $\Omega_1=50 \Gamma_{22},\Omega_2=10 \Gamma_{22}$,
$\Gamma_{33}=\Gamma_{22}$ and the detuning between the lasers is chosen to be
$\delta\omega=0.049\Gamma_{22}$. For the noise fields we choose $W_{33,max}= W_{44,max}$
and detunings as in Eqs. (\protect\ref{eq3},\protect\ref{eq4}).
The bandwidth of the noise fields are equal and chosen to be $6.66 \Gamma_{22}$.
The sharp rise to a
maximum for an intermediate value of the
noise intensity followed by a slower fall-off towards smaller SNR where
randomization dominates is a signature of the occurrence of SR.}
\end{center}
\end{figure}
we will discuss how the sensitivity for the detection
of a frequency mismatch between two coherent fields can be enhanced using incoherent
driving, i.e., employing SR.

Let us consider two coherent fields with Rabi frequencies $\Omega_1$ and
$\Omega_2$, whose oscillation frequencies differ by a small amount $\delta\omega$.
As it is well known, the superposition of two laser fields in a running wave
configuration results in a beating signal. The idea is to use this modulated signal as
our driving field (giving rise to the dressed states $|\pm\rangle$)
while tuning appropriately the additional incoherent driving
fields (see Fig. \ref{setup} for illustration). Numerical simulations
of the power spectrum for experimentally accessible parameters reveal again
that the frequency response of the atomic system exhibits stochastic resonance.
The signal to noise ratio for the first delta peak in the power spectrum is shown
in Fig. \ref{switch2}. The optimal performance, that is, the largest SNR, is
achieved for certain finite but non-zero noise intensity. This result implies
that stochastic resonance can have applications for example in frequency measurements.

Summarizing, we have showed that the phenomenon of SR, a paradigm of the
counter-intuitive role that noise may play in high sensitivity detection, can
be demonstrated at the level of a single ion. As an illustration of the
potential that this effect may have, we have discussed the use of a SR scheme
for the detection of the frequency difference of two lasers. As the proposed
experimental scenario relies on techniques well mastered by quantum opticians,
quantum SR may be expected to open up new experimental possibilities in
precision measurements at the quantum limit.

This work was supported by The Leverhulme Trust, the UK EPSRC and the European
Union. We would also like to thank D.M. Segal and R.C. Thompson and their
experimental ion trapping group at Imperial College for interesting
discussions and P.L. Knight for helpful comments on the manuscript.

\end{multicols}
\end{document}